\documentclass[prl,aps,12pt]{revtex4}
\usepackage{dsfont}
\usepackage{amsmath}
\usepackage{amssymb}
\usepackage{epsfig}
\usepackage{amscd}
\usepackage{ulem}

\begin{document}

\title{Implementation of controlled phase shift gates and Collins version of Deutsch-Jozsa algorithm on a quadrupolar 
spin-7/2 nucleus using non-adiabatic geometric phases} 
\author{T. Gopinath and Anil Kumar \\
{\small \it NMR Quantum Computing and Quantum Information Group.\\Department of Physics, and NMR Research Centre.\\
Indian Institute of Science, Bangalore - 560012, India.}}.

\begin{abstract}
In this work Controlled phase shift gates are implemented on a qaudrupolar system, by using 
non-adiabatic geometric phases. A general procedure is given, for implementing controlled 
phase shift gates in an 'N' level system. 
The utility of such controlled phase shift gates, is demonstrated here by implementing 
3-qubit Deutsch-Jozsa algorithm on a 7/2 quadrupolar nucleus oriented in a liquid crystal matrix.

\end{abstract}

\maketitle

\section{1. Introduction}

The use of quantum systems for information processing was first introduced by Benioff \cite{beni}. In 1985 Deutsch described quantum computers
which exploit the superposition of multi particle states, thereby achieving massive parallelism \cite{deu}.
Researchers have also studied the possibility of solving certain types of problems more efficiently than can be done on conventional computers
\cite{6deujoz,pw,gr}.
These theoretical possibilities have generated significant interest for experimental realization of quantum computers \cite{6ic,benson}.
Several techniques are being exploited for quantum computing and quantum information processing, including nuclear magnetic resonance (NMR) \cite{cor,ger}.

NMR has played a leading role for the practical demonstration of quantum gates and
algorithms.
Most of the NMR Quantum information processing (QIP) experiments have utilized systems having indirect spin-spin couplings (scalar J couplings)
\cite{nmr1,nmr2,nmr3,nmr5}. 
Recently NMR QIP has been demonstrated in quadrupolar and dipolar coupled systems, obtained by orienting the molecules
in liquid crystal media \cite{fung,6fung1}. In case of homo nuclear dipolar coupled systems, spins are often strongly coupled 
and hence can not be addressed individually \cite{mahesh}. 
However the $2^n$ eigen states are collectively treated as an n-qubit system \cite{fung,mahesh}.
Similarly for quadrupolar systems (spin $>$ 1/2), individual spins are treated as a multi-qubit system 
\cite{6fung1,6neeraj,6fung2,6fung3,6fung4,6mur,6pram2,6rana1,6bonk1,6bonk2,6bonk3,6rana2,6veeman}.
Resolved resonance
lines provide access to the full Hilbert space of $2^n$ dimension \cite{6fung1,6neeraj}.

A quadrupolar nucleus of spin I has (2I+1) energy levels, which are equi-spaced in the high magnetic field and the 2I single quantum 
transitions are degenerate. In the presence of first order quadrupolar interaction this degeneracy is lifted and gives rise to 2I 
equi-spaced transitions \cite{6ern,6dhiel}. For molecules partially oriented 
in anisotropic media, such as liquid crystals, such a situation is often obtained \cite{6fung1}. If $(2I+1)=2^n$, such a system can 
be treated as an n-qubit system \cite{6fung1,6neeraj}. The advantage of such systems and the dipolar coupled systems, over 
the J-coupled systems 
is that the coupling values are one to two orders of magnitude larger, allowing shorter gate times or the use of shorter transition 
selective 
pulses. So far quadrupolar systems have been used for, quantum simulation, preparation of pseudo pure states, implementation of 
quantum gates and search algorithms \cite{6fung2,6fung3,6fung4,6mur,6pram2,6rana1,6bonk1,6bonk2,6bonk3,6veeman}. 
Recently Das et al have implemented Cleve version of 2-qubit DJ algorithm on a spin 7/2 nucleus \cite{6rana2}.
In all these cases the 
controlled gates are 
implemented by inverting populations 
between various levels, by using transition selective $\pi$ pulses. 
Recently it has been demonstrated that non-adiabatic geometric phases can be used for implementing quantum gates \cite{6rana3,6gopi}. 
Here we use non-adiabatic geometric phases to implement controlled phase shift gates and Collins version of DJ 
algorithm on a 3-qubit system obtained by eight eigen states of a spin-7/2 quadrupolar nucleus.

The Hamiltonian of a quadrupolar nucleus partially oriented in liquid crystal matrix, in the presence of large magnetic 
field $B_o$, having a first order quadrupolar coupling, is given by,

\begin{eqnarray}
H=H_z+H_Q=\omega_o I_z+\frac{2\pi e^2qQ}{[4I(2I-1)]}S(3I_z^2-I^2) =\omega_o I_z+2\pi \wedge (3I_z^2-I^2),  \label{6eqn1} 
\end{eqnarray}

where $\omega_o=-\gamma B_o$ is the resonance frequency, S is the order parameter 
at the site of the nucleus, and $e^2qQ$ is the quadrupolar coupling. If the order parameter S is small, the effective 
quadrupolar coupling $\wedge$ can be of the order of a few kHz in spite of $e^2qQ$ being of the order of several MHz. 
Thus, it is possible to observe the satellite transitions due to first order quadrupolar coupling.

\begin{figure}
\begin{center}
\epsfig{file=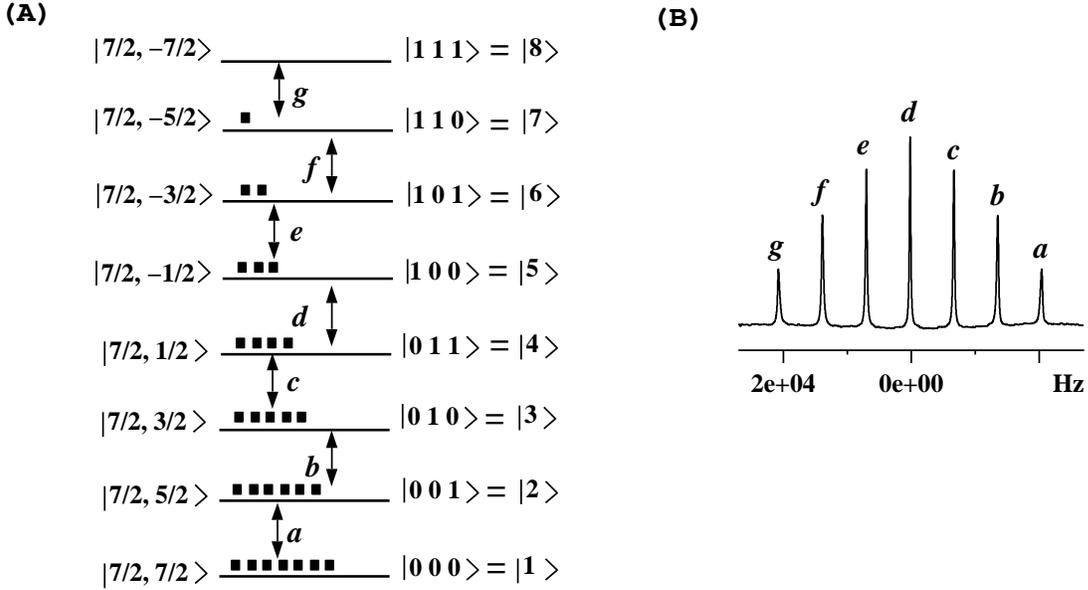,width=8cm,angle=270}
\caption{(A) Schematic energy level diagram of a spin-7/2 nucleus, spin states $\vert 7/2, 7/2 \rangle$ ....$\vert 7/2, -7/2 \rangle$ are
labeled as eight basis states of a three qubit system, the dark rectangles represent the equilibrium populations, and the single quantum
transitions are labeled as a, b...., g. (B) Equilibrium spectrum of oriented $^{133}Cs$ nucleus (spin-7/2), obtained by a
non-selective $(\pi/2)_y$ pulse, at a resonance frequency of 65.59 MHz on a Bruker AV-500 spectrometer at a temperature $307K$. The
distance between successive transitions is equal to the effective quadrupolar coupling ($\wedge$).
The line widths of transitions a, b, ...,g are observed to be in the ratio
3.2:2.3:1.5:1.0:1.5:2.2:3.1
and the integrated experimental intensities are in the ratio
7.2:12.1:15.1:16:15.1:12.0:7.0 (theoretically expected 7:12:15:16:15:12:7)}
\label{equi}
\end{center}
\end{figure}

A 50-50 mixture of Cesium-pentadecafluoro-octanate and $D_2O$ forms a lyotropic liquid crystal at room temperature \cite{6fung2}.
The schematic energy level diagram of oriented spin-7/2 nucleus is shown in Fig. (\ref{equi}A). The 8 energy levels are labeled as
the basis states of a three qubit system.
In the high field approximation, effective quadrupolar coupling ($\wedge$) can be considered as a small perturbation to Zeeman field.
Thus for population purposes the equilibrium density matrix can be considered to be proportional to $H_z$ (Fig. \ref{equi}A). 
Partially oriented $^{133}{Cesium}$ nucleus (I=7/2) gives rise to a well resolved 7 transitions spectrum at room temperatures ranging 
from 290 K to 315 K Fig (\ref{equi}B).
The effective quadrupolar coupling ($\wedge$) changes with temperature,
since the order parameter is a function of temperature. All the experiments
have been carried out here at temperature 307 K, which gives $\wedge = 6856$ Hz.
The equilibrium spectrum is obtained by applying a
$(\pi/2)_y$ pulse, is shown in Fig.(\ref{equi}B). 
The integrated intensities are in the expected ratio 
7:12:15:16:15:12:7, as determined by the transition matrix elements of $\vert (I_{x(y)})_{ij} \vert ^2$ (Appendix) \cite{6ern}.

\section{2. Controlled phase shift gates}

Non-adiabatic geometric phase in NMR was first verified by Suter et.al \cite{6pines1}.
Non adiabatic geometric phases in NMR, were used to implement controlled phase shift gates, Deutsch-Jozsa (DJ) and Grover algorithms in 
weakly J-coupled and strongly dipolar coupled systems \cite{6rana3,6gopi}. 

\begin{figure}
\begin{center}
\epsfig{file=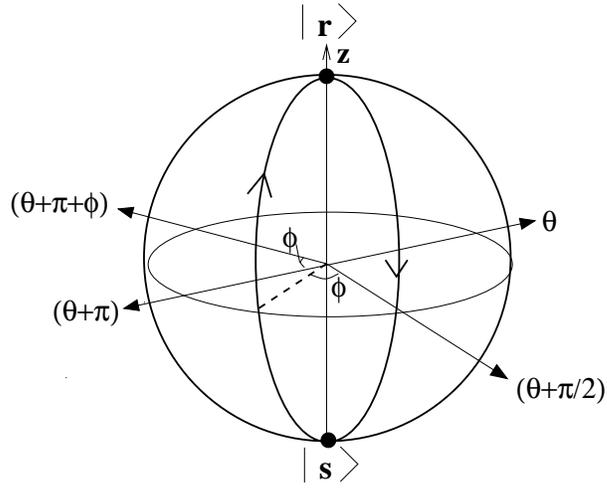,width=8cm,angle=0}
\caption{A two level subspace (r,s) is represented on a Bloch sphere. Two
$\pi$ pulses applied on a transition (r,s) with phases $\theta$ and $(\theta+\pi+\phi)$, cyclically rotates the states 
$\vert r \rangle$ and $\vert s \rangle$. Solid angle subtended by this closed loop at the center
of the sphere is, $2\phi$. The phase acquired by the states
$\vert r \rangle$ and $\vert s \rangle$ is $e^{i\phi}$ and $e^{-i\phi}$ respectively.}  \label{quad-bloch}
\end{center}
\end{figure}

A two level subspace (r,s) forms a fictitious spin-1/2 subspace, hence the states $\vert r \rangle$ and 
$\vert s \rangle$ can be represented on a Bloch sphere (Fig. \ref{quad-bloch}) \cite{6gopi}. 
The two $\pi$ pulses applied on a transition (r,s), cyclically rotates each of the states 
$\vert r \rangle$ and $\vert s \rangle$ \cite{6gopi}. The cyclic path followed by each of the two states, makes an equal 
solid angle ($\Omega$) at the center of the 
Bloch sphere \cite{6gopi}. The geometric phase acquired by the states $\vert r \rangle$ and $\vert s \rangle$ is given by 
$e^{i\Omega/2}$ and $e^{-i\Omega/2}$ respectively \cite{6rana3,6gopi}, where the phase difference of $e^{i\pi}$ between the states indicate 
that the two states are traversed in the opposite direction \cite{6pines1,6rana3}. 
It is possible to obtain various solid angles, hence various phases, by shifting the phases of the two $\pi$ pulses with 
respect to each other (Fig. \ref{quad-bloch}) \cite{6rana3,6gopi}. 
It is shown that two $\pi$ pulses with phases $\theta$ and $(\theta+\pi+\phi)$ respectively, applied 
on a transition (r,s), subtends a solid angle $\Omega=2\phi$, thus the phase acquired by the states 
$\vert r \rangle$ and $\vert s \rangle$ is $e^{i\phi}$ and $e^{-i\phi}$ respectively.

However the internal Hamiltonian evolves during the application of a transition selective $\pi$ pulse, and gives rise to an 
additional phase 
known as dynamical phase. In order to observe only the geometric phase, one has to refocus the evolution of 
internal Hamiltonian. As shown in Eq. (\ref{6eqn1}), the Hamiltonian consists of Larmor frequency term and quadrupolar coupling term. 
The Larmor frequency is identical to the frequency of central transition 'd'. Thus the receiver frequency is set at frequency of 
transition 'd' (Fig. \ref{equi}A), hence there is no evolution due to the Larmor frequency term. The duration of the selective 
$\pi$ pulse is chosen in such a way that the phase acquired by the 
quadrupolar term makes a complete $2\pi$ rotation, in other words 
$e^{-iH_Qt_p}=1$, where $t_p$ is duration of the selective $\pi$ pulse which is equal to 1.425 ms.

\begin{figure}
\begin{center}
\epsfig{file=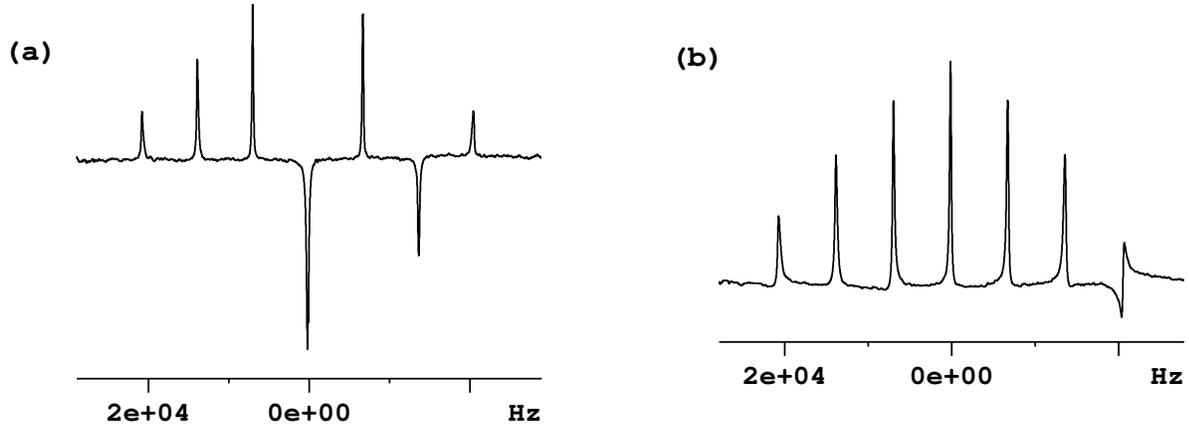,width=6cm,angle=270}
\caption{Implementation of controlled phase shift gates (a) $U^{\pi}_{(3,4)}=diag[1, 1, -1, -1, 1, 1, 1, 1]$ and
(b) $U^{\pi/2}_{1}=diag[e^{i\pi/2}, 1, 1, 1, 1, 1, 1, 1]$, preceded by a hard $(\pi/2)_y$ pulse. $U^{\pi}_{(3,4)}$ is implemented by
applying two $(\pi)_y$ pulses on transition c. $U^{\pi/2}_{1}$ is implemented by applying four multi-frequency $\pi$
pulses (Eq. \ref{6eqn5}).}  \label{gates}
\end{center}
\end{figure}

The phase information can be encoded in the coherences, hence in order to observe the phases the initial state should 
contain the coherences, which is obtained by applying a non-selective $(\pi/2)_y$ pulse on equilibrium state. Figure (\ref{gates}a) 
shows the implementation of controlled phase shift gate, represented by the diagonal matrix 
$U_{(3,4)}^{\pi}=diag[1, \hspace{0.2cm} 1, \hspace{0.2cm} -1, \hspace{0.2cm} -1, \hspace{0.2cm} 1, \hspace{0.2cm} 1,
\hspace{0.2cm} 1,\hspace{0.2cm} 1]$, obtained by applying two selective $\pi$ pulses on transition (3,4), with phases y and 
$(y+\pi+\pi)=y$. The two states 
$\vert 010 \rangle$ and $\vert 011 \rangle$ acquire $\pi$ phase shift so the transitions 'b' and 'd' are inverted, confirming the phase gate. 
since the phases of two $\pi$ pulses are same, it is possible to combine the two $\pi$ pulses in to a 
single $2\pi$ pulse. Such pulses are used in section (3.1). 

In some quantum computing protocols 
(for example Grover search algorithm and quantum Fourier transform) one requires a phase gate such as $U_k^{\phi}$ (Table. 1), 
which creates the relative phase $e^{i\phi}$ at $k^{th}$ state, for example $U_8^{\phi}$=$[1, 1, 1, 1, 1, 1, 1, e^{i\phi}]$. 
Such phase gates can be implemented by sandwiching 
various phase gates $u_{(i,j)}^{\phi}$, as shown in Table (1). 
$u_{(i,j)}^{\phi}$ indicates the geometric phase shift gate obtained by 
applying a pair of selective $\pi$ pulses on a transition (i,j) with phases y and $(y+\pi+\phi)$, thus the diagonal elements 
corresponding to i and j states, acquire 
a phase shift 
$e^{i\phi}$ and $e^{-i\phi}$ respectively. For example, 
$u_{(1,2)}^{\phi/8}$=diag[$e^{i\phi/8}$, $e^{-i\phi/8}$, 1, 1, 1, 1, 1, 1], which can be obtained by applying two selective $\pi$ pulse on 
transition (1,2) with phases y and $(y+\pi+\phi/8)$.

\begin{figure}
\begin{center}
\epsfig{file=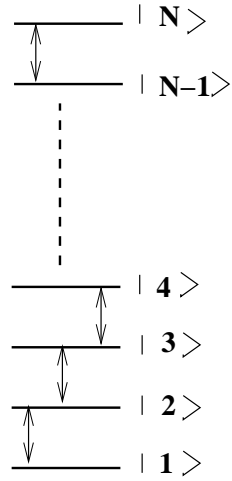,width=3cm,angle=0}
\caption{Schematic energy level diagram of an N-level system. The arrows represent single quantum transitions.}    \label{nlevels}
\end{center}
\end{figure}

The method of constructing controlled phase shift gates (Table 1),
is explained here. For example, consider a system consisting
of N energy levels and (N-1) single quantum transitions between the levels (1,2), (2,3),.....(N-1,N), as shown in Fig.(\ref{nlevels}).
In this
system, the controlled phase shift gate
$U_k^{\phi}= diag[1_1, \hspace{0.1cm} 1_2, ......,\hspace{0.1cm} (e^{i\phi})_k, \hspace{0.1cm} ...... ,\hspace{0.1cm} 1_N]$, can be implemented
by sandwiching (N-1) phase shift gates, as,

\begin{eqnarray}
U_k^{\phi}= u_{(1,2)}^{-\phi/N}.u_{(2,3)}^{-2\phi/N}.-----.u_{(k-1,k)}^{-(k-1)\phi/N}.u_{(k,k+1)}^{(N-k).\phi/N}.u_{(k+1,k+2)}^{(N-(k+1)).
\phi/N}.
------.u_{(N-1,N)}^{\phi/N}. \label{eqn5}
\end{eqnarray}

In Eq. (\ref{eqn5}), by putting N=8 one can obtain any of the $U_k^{\phi}$ shown in Table (1).

In Table (1), each of the operators, $U_k^{\phi}$, requires seven pairs of selective $\pi$ pulses, where each pair of 
pulses will act on 
a single transition. 
However the $\pi$ pulses on unconnected transitions can be applied simultaneously. Hence one can 
simultaneously implement $u_{(1,2)}$, $u_{(3,4)}$, $u_{(5,6)}$, $u_{(7,8)}$ by using a pair of 
multi-frequency (MF) pulses, similarly $u_{(2,3)}$, $u_{(4,5)}$, $u_{(6,7)}$ can be implemented by using another pair of 
multi-frequency pulses. 
For example $U_1^{\phi}$ can be implemented by using 
 seven geometric phase shift gates (Table 1), obtained by using two pairs of multi-frequency $\pi$ pulse, as described below. 

\begin{eqnarray}
U_{1}^\phi&=&u_{(1,2)}^{7\phi/8}.u_{(2,3)}^{3\phi/4}.u_{(3,4)}^{5\phi/8}.u_{(4,5)}^{\phi/2}.
u_{(5,6)}^{3\phi/8}.u_{(6,7)}^{\phi/4}.u_{(7,8)}^{\phi/8}
\label{6eqn13}
\end{eqnarray}

The unitary operators of phase shift gates correspond to diagonal matrices which commute with each other, hence $U_1^{\phi}$ of 
Eq. (\ref{6eqn13}) can be rewritten as,

\begin{eqnarray}
U_{1}^\phi&=& \{[u_{(1,2)}^{7\phi/8}].[u_{(3,4)}^{5\phi/8}].[u_{(5,6)}^{3\phi/8}].[u_{(7,8)}^{\phi/8}]\}.\{[u_{(2,3)}^{3\phi/4}].
[u_{(4,5)}^{\phi/2}].[u_{(6,7)}^{\phi/4}]\} \nonumber \\
&=& \{ [(\pi)_y^a(\pi)_{y+\pi+7\phi/8}^a].[(\pi)_y^c(\pi)_{y+\pi+5\phi/8}^c].[(\pi)_y^e(\pi)_{y+\pi+3\phi/8}^e].
[(\pi)_y^g(\pi)_{y+\pi+\phi/8}^g]\}. 
\nonumber \\
&&\{ [(\pi)_y^b(\pi)_{y+\pi+3\phi/4}^b].[(\pi)_y^d(\pi)_{y+\pi+\phi/2}^d].[(\pi)_y^f(\pi)_{y+\pi+\phi/4}^f]\} \nonumber \\
&=& \{ [(\pi)^{a,c,e,g}_y.(\pi)^{a,c,e,g}_{(y+\pi+7\phi/8),(y+\pi+5\phi/8),(y+\pi+3\phi/8),(y+\pi+\phi/8)}]   \}.\nonumber \\ 
&& \{ [(\pi)^{b,d,f}_y.(\pi)^{b,d,f}_{(y+\pi+3\phi/4),(y+\pi+\phi/2),(y+\pi+\phi/4)}]   \}   \label{6eqn5}
\end{eqnarray}

The $\pi$ pulses acting on unconnected transitions, a, c, e and g (Eq. \ref{6eqn5}), can be combined (added) in to a single pulse, since the 
operators of the 
subspaces corresponding to these transitions, $I_{x(y)}^{(1,2)}$, $I_{x(y)}^{(3,4)}$, $I_{x(y)}^{(5.6)}$ and $I_{x(y)}^{(7,8)}$ \cite{6gopi}, 
commute with each other. Similarly one can combine the $(\pi)$ pulses acting unconnected transitions b, d and f (Eq. \ref{6eqn5}). 
Figure (\ref{gates}b) shows the spectrum obtained by a $(\pi/2)_y$ pulse followed by the implementation of $U_1^{\phi}(\phi=\pi/2)$ using 
four MF-$\pi$ pulses (Eq. \ref{6eqn5}), the state $\vert 000 \rangle$ hence the transition 'a' acquires a $\pi/2$ phase shift, 
confirms $U_1^{\pi/2}$.

\begin{table}
\begin{center}
 Table (1). Unitary operators of controlled-$\phi$ phase shifted gates \\
\begin{tabular}
{|c|} \hline
$U_{k}^\phi$ $(k= 1,2, ----,8)$ \\ \hline
$U_{1}^\phi=diag[e^{i\phi},1,1,1,1,1,1,1]=u_{(1,2)}^{7\phi/8}.u_{(2,3)}^{3\phi/4}.u_{(3,4)}^{5\phi/8}.u_{(4,5)}^{\phi/2}.
u_{(5,6)}^{3\phi/8}.u_{(6,7)}^{\phi/4}.u_{(7,8)}^{\phi/8}$  \\ \hline
$U_{2}^\phi=diag[1,e^{i\phi},1,1,1,1,1,1]=u_{(1,2)}^{-\phi/8}.u_{(2,3)}^{3\phi/4}.u_{(3,4)}^{5\phi/8}.u_{(4,5)}^{\phi/2}.u_{(5,6)}^
{3\phi/8}.u_{(6,7)}^{\phi/4}.u_{(7,8)}^{\phi/8}$  \\ \hline
$U_{3}^\phi=diag[1,1,e^{i\phi},1,1,1,1,1]=u_{(1,2)}^{-\phi/8}.u_{(2,3)}^{-\phi/4}.u_{(3,4)}^{5\phi/8}.u_{(4,5)}^{\phi/2}.u_{(5,6)}^
{3\phi/8}.u_{(6,7)}^{\phi/4}.u_{(7,8)}^{\phi/8}$  \\ \hline
$U_{4}^\phi=diag[1,1,1,e^{i\phi},1,1,1,1]=u_{(1,2)}^{-\phi/8}.u_{(2,3)}^{-\phi/4}.u_{(3,4)}^{-3\phi/8}.u_{(4,5)}^{\phi/2}.u_{(5,6)}^
{3\phi/8}.u_{(6,7)}^{\phi/4}.u_{(7,8)}^{\phi/8}$  \\  \hline
$U_{5}^\phi=diag[1,1,1,1,e^{i\phi},1,1,1]=u_{(1,2)}^{-\phi/8}.u_{(2,3)}^{-\phi/4}.u_{(3,4)}^{-3\phi/8}.u_{(4,5)}^{\phi/2}.u_{(5,6)}^
{3\phi/8}.u_{(6,7)}^{\phi/4}.u_{(7,8)}^{\phi/8}$  \\ \hline
$U_{6}^\phi=diag[1,1,1,1,1,e^{-i\phi},1,1]=u_{(1,2)}^{-\phi/8}.u_{(2,3)}^{-\phi/4}.u_{(3,4)}^{-3\phi/8}.u_{(4,5)}^{-\phi/4}.u_{(5,6)}^
{-5\phi/8}.u_{(6,7)}^{\phi/4}.u_{(7,8)}^{\phi/8}$  \\ \hline
$U_{7}^\phi=diag[1,1,1,1,1,1,e^{i\phi},1]=u_{(1,2)}^{-\phi/8}.u_{(2,3)}^{-\phi/4}.u_{(3,4)}^{-3\phi/8}.u_{(4,5)}^{-\phi/2}.u_{(5,6)}^
{-5\phi/8}.
u_{(6,7)}^{-6\phi/8}.u_{(7,8)}^{\phi/8}$ \\ \hline
$U_{8}^\phi=diag[1,1,1,1,1,1,1,e^{-i\phi}]=u_{(1,2)}^{-\phi/8}.u_{(2,3)}^{-\phi/4}.u_{(3,4)}^{-3\phi/8}.u_{(4,5)}^{-\phi/2}.u_{(5,6)}^
{-5\phi/8}.u_{(6,7)}^{-3\phi/4}.u_{(7,8)}^{-7\phi/8}$  \\ \hline
\end{tabular}
\end{center}
\end{table}

\section{3. Collins version of DJ algorithm}
Deutsch-Jozsa (DJ) algorithm determines in a single query, whether a given function is constant or 
balanced \cite{6deujoz,6cleve}. A function
is "constant" if it gives the same
output for all inputs, and "balanced" if it gives one output for half the number of inputs and another for the remaining 
half \cite{6ic}. 
Classically for an n bit binary function, at least ($2^{n-1}+1$) queries are needed to determine whether the function is constant or
balanced, whereas the DJ algorithm requires only a single query \cite{6ic}. The Cleve version of the DJ algorithm requires
 an extra qubit
(ancilla qubit) and uses controlled-not gates \cite{6cleve}, whereas Collins version does not require the ancilla qubit but needs 
controlled phase shift gates \cite{6col}.

\begin{figure}
\begin{center}
\epsfig{file=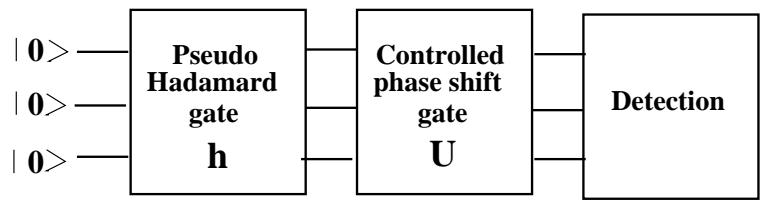,width=10cm,angle=0}
\caption{Quantum circuit for Collins version of DJ algorithm on a three qubit system.}    \label{djckt}
\end{center}
\end{figure}

For a three qubit system there are
2 constant and 70 ($c^N_{N/2}$=$c^8_4$) balanced functions \cite{6col,6mangold}. The quantum circuit of the Collins version of the DJ algorithm is shown in 
Fig. (\ref{djckt}). The
algorithm starts with a pure state (pseudo pure state in NMR) $\vert 000 \rangle$ which is converted 
to a superposition
state of eight basis states $\vert 000 \rangle$, $\vert 001 \rangle$, .....,$\vert 111 \rangle$, by applying a
pseudo Hadamard gate on all 
the three qubits. Thereafter a unitary operator $U$ (oracle) is applied,
followed by detection. Theoretically, after
$U$ one
has to apply the Hadamard gate. In NMR, a Hadamard gate is replaced by a pseudo Hadamard gate, which is implemented by a
$(\pi/2)_{y}$ pulse, and for detection another $(\pi/2)_{-y}$ pulse is needed. These two pulses cancel each other and the result of the algorithm is
available immediately
after $U$.

The unitary operators (U) of the Oracle, are eight dimensional diagonal matrices. 
For two constant functions, U's are given by, 

\begin{eqnarray}
&U_{c1}&=I \text{  (Unit matrix) and} \nonumber \\
&U_{c2}&=diag[-1, -1, -1, -1, -1, -1, -1, -1]=(-1).I.  
\end{eqnarray}

For balanced functions, we define the unitary operator as $U(1,k,l,m)$, which means that 
each of the four diagonal elements 1,k,l, and m are equal to -1, whereas the remaining four diagonal elements are 1.
For example $U(1,2,3,4)=diag[-1,-1,-1,-1,1,1,1,1]$. The unitary operators of 35 balanced functions can be written as, 

\begin{eqnarray}
U(1,k,l,m)=[-\vert 1 \rangle \langle 1 \vert - \vert k \rangle \langle k \vert - \vert l \rangle \langle l \vert 
- \vert m \rangle \langle m \vert].I, \nonumber \\
where \hspace{0.2cm} k<l<m, \hspace{0.2cm} k=2,3..,6, \hspace{0.2cm} l \ge k+1, \hspace{0.2cm} and \hspace{0.2cm} m \ge l+1   \label{6eqn6}
\end{eqnarray}

The unitary operators of the remaining 35 balanced functions are identical to one of the operators 
of Eq. (\ref{6eqn6}), up to an overall phase factor $e^{i\pi}$ \cite{6mangold}.
Here we implement the unitary operators corresponding to 11 of the 35 balanced functions (Eq. \ref{6eqn6}) and one constant 
function ($U_{c1}$).

\subsection{3.1. Experimental Implementation}

\begin{figure}
\begin{center}
\epsfig{file=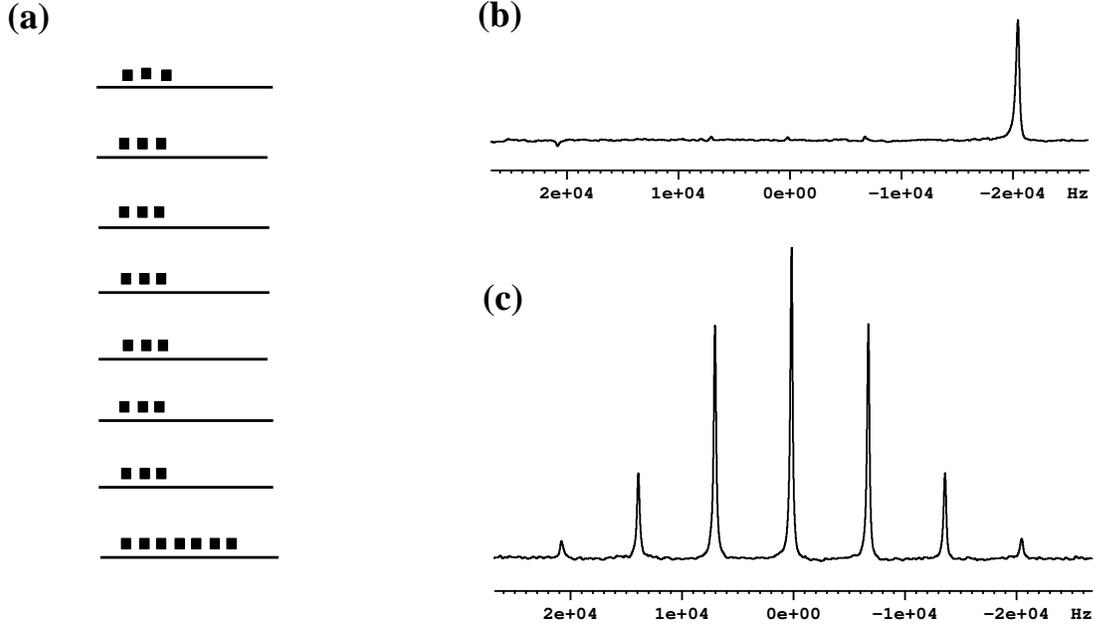,width=10cm,angle=270}
\caption{(a) Population distribution of $\vert 000 \rangle$ PPS obtained by applying an amplitude modulated multi-frequency pulse
(explained in text) on transitions b, c, d, e, f, and g at equilibrium state. (b) Spectrum of $\vert 000 \rangle$ PPS obtained by
applying a non selective $5^o$
pulse with 32 scans. (c) Coherent superposition state (Eq.s \ref{6eqn7},\ref{6eqn8}) obtained by applying a $(\pi/2)_y$ pulse
on $\vert 000 \rangle$ PPS. The intensities are in accordance with Eq. (\ref{6eqn7}), which is different from equilibrium
intensities of Fig. \ref{equi}. Experimental integrated intensities are in the ratio 
10:48:101:140:100:46:8, while the expected theoretical intensities are in the ratio 7:42:105:140:105:42:7. The intensities here 
represent unequal superposition of basis states.}
\label{pps}
\end{center}
\end{figure}

{\bf (i)Preparation of pseudo pure state (PPS) $\vert000\rangle$:}
A multi frequency pulse with six harmonics
 with appropriate amplitudes was applied. The six harmonics are the frequencies of the six leftmost transitions b,c,..., g of 
Fig. (\ref{equi}B). 
Under the influence of this pulse, the populations of all the states except $\vert 000\rangle$, start a collective population
transfer among them as a linear chain of coupled oscillators \cite{6fung2}. 
Choosing the correct amplitudes of various harmonics and duration of the pulse, these six transitions can be simultaneously saturated 
with average population of these seven levels. 
The relative amplitudes of the 
six harmonics b, c, ..., and g are respectively given by 0.84, 0.93, 1, 1.03, 1.04 and 1.07 \cite{6rana2}. 
The duration of the pulse is 2.05 ms. A gradient pulse was applied subsequently to kill (dephase) the coherences created during the
process. The final populations were measured using a non-selective small-angle (5$^o$) pulse.
The small-angle pulse, converts the population differences between adjacent energy levels into single quantum transitions,
within linear approximation. The spectrum of Fig (\ref{pps}b) confirms the preparation of $\vert 000\rangle$ pseudopure state.

{\bf (iii) Coherent Superposition:} After the creation of $\vert000\rangle$ PPS, a pseudo Hadamard gate (hard $(\pi/2)_{y}$ pulse)
creates a superposition of 8 basis states. It may be noted that unlike in weakly coupled spins, the state created here is not in 
equal superposition
of eigenstates, since the coefficients of various eigenstates are different. However, as is shown here, the created coherent
superposition can be
utilized for quantum parallelism, to distinguish
different classes of functions. The state of the system after a $(\pi/2)_{y}$ pulse on $\vert000\rangle$ PPS, is given by, 

\begin{eqnarray}
\vert \psi \rangle &=& (e^{-i\frac{\pi}{2}I_y}). \vert 000 \rangle  \nonumber \\ 
&=&\frac{1}{8\sqrt{2}}[\vert 000 \rangle + \sqrt{7} \vert 001 \rangle + \sqrt{21} \vert 010 \rangle 
+ \sqrt{35} \vert 011 \rangle + \sqrt{35} \vert 100 \rangle \nonumber \\ 
&&+ \sqrt{21} \vert 101 \rangle + \sqrt{7}\vert 110 \rangle + \vert 111 \rangle] \nonumber \nonumber \\ 
&=&\frac{1}{8\sqrt{2}}[\vert 1 \rangle + \sqrt{7} \vert 2 \rangle + \sqrt{21} \vert 3 \rangle + \sqrt{35} \vert 4 \rangle +
\sqrt{35} \vert 5 \rangle + \sqrt{21} \vert 6 \rangle + \sqrt{7} \vert 7 \rangle + \vert 8 \rangle].     \label{6eqn7}
\end{eqnarray}

The corresponding density matrix can be written as,

\begin{eqnarray}
&& \begin{matrix} ~~~~\vert 1 \rangle& ~~~~~~ \vert 2 \rangle&~~~~~~~~ \vert 3 \rangle& ~~~~~~~ \vert 4 \rangle&~~~~~~ \vert 5 \rangle
&~~~~~~ \vert 6 \rangle& ~~~~~~~ \vert 7 \rangle&~~~~~ \vert 8 \rangle \end{matrix} \nonumber \\ 
\sigma= \vert \psi \rangle \langle \psi \vert =\frac{1}{128}. 
&&\begin{pmatrix}
1&~~\fbox{$\sqrt{7}$}&~~\sqrt{21}&~~\sqrt{35}&~\sqrt{35}&~\sqrt{21}&~~\sqrt{7}&~1 \cr
\fbox{$\sqrt{7}$}&~~7&~~\fbox{$\sqrt{147}$}&~~\sqrt{245}&~\sqrt{245}&~\sqrt{147}&~~7&~\sqrt{7} \cr
\sqrt{21}&~~\fbox{$\sqrt{147}$}&~~21&~~\fbox{$\sqrt{735}$}&~\sqrt{735}&~21&~~\sqrt{147}&~\sqrt{21} \cr
\sqrt{35}&~~\sqrt{245}&~~\fbox{$\sqrt{735}$}&~~35&~\fbox{35}&~\sqrt{245}&~~\sqrt{147}&~\sqrt{35} \cr
\sqrt{35}&~~\sqrt{245}&~~\sqrt{735}&~~\fbox{35}&~35&~\fbox{$\sqrt{735}$}&~~\sqrt{147}&~\sqrt{35} \cr
\sqrt{21}&~~\sqrt{147}&~~21&~~\sqrt{735}&~\fbox{$\sqrt{735}$}&~21&~~\fbox{$\sqrt{147}$}&~\sqrt{21} \cr
\sqrt{7}&~~7&~~\sqrt{147}&~~\sqrt{245}&~\sqrt{245}&~\fbox{$\sqrt{147}$}&~~7&~\fbox{$\sqrt{7}$} \cr
1&~~\sqrt{7}&~~\sqrt{21}&~~\sqrt{35}&~\sqrt{35}&~\sqrt{21}&~~\fbox{$\sqrt{7}$}&~1 \cr
\end{pmatrix} \nonumber 
\begin{matrix}\vert 1 \rangle \cr \vert 2 \rangle \cr \vert 3 \rangle \cr \vert 4 \rangle \cr \vert 5 \rangle
\cr \vert 6 \rangle \cr \vert 7 \rangle \cr \vert 8 \rangle \end{matrix}. \\ \label{6eqn8}
\end{eqnarray}

where the elements within the boxes represent single quantum transitions a, b,.., g. Figure (\ref{pps}c) shows the spectrum of 
$\vert \psi \rangle$, where the intensities of transitions are obtained by modulus of the product of  
single quantum elements of Eq. \ref{6eqn8}, with corresponding matrix elements of $I_x$.

{\bf (iv) Implementation of U:}
For constant function $U=U_{c1}=I$ (Unit matrix), which requires no pulse, thus the final state $\psi_{c1}$ is given by, 
\begin{eqnarray}
\vert \psi_{c1} \rangle= U_{c1} \vert \psi \rangle= \vert \psi \rangle    \label{6eqn9}
\end{eqnarray}

As mentioned in section (2), the controlled phase shift gate $U_{(i,j)}^{\pi}$ can be achieved by applying 
two $(\pi)$ pulses with same phase on transition (i,j), which can be be combined in to a single $(2\pi)$ pulse. 
For balance functions, U = $U(1,k,l,m)$ (Eq. \ref{6eqn6}), can be decomposed in to, 

\begin{eqnarray}
U(1,k,l,m)&=&u_{(1,k)}^{\pi}.u_{(l,m)}^{\pi}. \nonumber \\
&=&[u_{(1,2)}^{\pi}....u_{(k-1,k)}^{\pi}][u_{(l,l+1)}^{\pi}.u_{(l+1,l+2)}^{\pi}...u_{(m-1,m)}^{\pi}]
\label{6eqn11}
\end{eqnarray}

In Eq. (\ref{6eqn11}), $u_{(1,k)}^{\pi}$ and $u_{(l,m)}^{\pi}$ are implemented by applying two $(2\pi)_y$ pulses on transitions 
(1,k) and (l,m) respectively, if (1,k) and (l,m) does not correspond to single quantum transitions, then they can be decomposed 
in to a series of single quantum transitions, as shown in the second equality of Eq. (\ref{6eqn11}), for example, 

\begin{eqnarray}
U(1,2,3,4)=u_{(1,2)}^{\pi}.u_{(3,4)}^{\pi}&=&(2\pi)^a.(2\pi)^c, \nonumber \\
U(1,4,5,8)=u_{(1,4)}^{\pi}.u_{(5,8)}^{\pi}&=&[u_{(1,2)}^{\pi}.u_{(2,3)}^{\pi}.u_{(3,4)}^{\pi}].[u_{(5,6)}^{\pi}.
u_{(6,7)}^{\pi}.u_{(7,8)}^{\pi}] \nonumber \\
&=&[(2\pi)^a.(2\pi)^b.(2\pi)^c].[(2\pi)^e.(2\pi)^f.(2\pi)^g] \nonumber \\
&=&(2\pi)^{a,c,e,g}.(2\pi)^{b,f},
\label{6eqn12}
\end{eqnarray}

where in the last equality of Eq. (\ref{6eqn12}), the pulses acting on unconnected transitions (Fig. \ref{equi}) are combined in to a single MF pulse. 
The required pulses for various $U(1,k,l,m)$ of Eq. (\ref{6eqn6}) are given in Table (2), 
the $2\pi$ pulses acting on unconnected transitions, are combined in to a single multi-frequency (MF) $2\pi$ pulse.
The duration of each transition selective pulse, is set such that
the evolution due to quadrupolar coupling, makes a complete $2\pi$ rotation.
It is to be noted that, in Eq. (\ref{6eqn12}) and Table (2), one can use
any phase for ($2\pi$) pulses, we have used 'y' phase for all the pulses. 
The final state $\vert \psi_{1,k,l,m} \rangle$ is given by

\begin{eqnarray}
\vert \psi_{1,k,l,m} \rangle= U(1,k,l,m)\vert \psi \rangle.   \label{6eqn10}
\end{eqnarray} 

It can be seen that $\vert \psi_{1,k,l,m} \rangle$ is same as $\vert \psi \rangle$, except that the basis states $\vert 1 \rangle$, $\vert k \rangle$, 
$\vert l \rangle$ and $\vert m \rangle$ 
acquire a phase factor $e^{i\pi}$ (=-1).

\begin{table}[h!]
\begin{center}
Table (2). Unitary operators of balanced functions (Eq. \ref{6eqn6}) and corresponding pulse sequences, consisting of $2\pi$ pulses, where 
all the pulses are applied with phase 'y'.
\hspace*{-0.85cm}
\begin{tabular}
{|c|c||c|c||c|c|} \hline \hline
   U(1,k,l,m) & Pulse sequence & U(1,k,l,m) & Pulse sequence & U(1,k,l,m) & Pulse sequence  \\ \hline \hline
 U(1,2,3,4)  & $(2\pi)^{a,c}$  & U(1,2,3,5)  & $(2\pi)^{a,c}$-$(2\pi)^{d}$  & U(1,2,3,6)  & $(2\pi)^{a,c,e}$-$(2\pi)^{d}$ \\ \hline
 U(1,2,3,7)  & $(2\pi)^{a,c,e}$- $(2\pi)^{d,f}$ & U(1,2,3,8)  & $(2\pi)^{a,c,e,g}$-$(2\pi)^{d,f}$ & U(1,2,4,5)  & $(2\pi)^{a,d}$  \\ \hline
 U(1,2,4,6)  & $(2\pi)^{a,d}$-$(2\pi)^{e}$  & U(1,2,4,7)  & $(2\pi)^{a,d,f}$-$(2\pi)^{e}$ & U(1,2,4,8)  & $(2\pi)^{a,d,f}$-$(2\pi)^{e,g}$
\\ \hline
 U(1,2,5,6)  & $(2\pi)^{a,e}$  & U(1,2,5,7)  & $(2\pi)^{a,e}$-$(2\pi)^{f}$ & U(1,2,5,8)  & $(2\pi)^{a,e,g}$-$(2\pi)^{f}$  \\ \hline
 U(1,2,6,7)  & $(2\pi)^{a,f}$  & U(1,2,6,8)  & $(2\pi)^{a,f}$-$(2\pi)^{g}$ & U(1,2,7,8)  & $(2\pi)^{a,g}$  \\ \hline
 U(1,3,4,5)  & $(2\pi)^{a,d}$-$(2\pi)^{b}$  & U(1,3,4,6)  & $(2\pi)^{a,d}$-$(2\pi)^{b,e}$ & U(1,3,4,7)  & $(2\pi)^{a,d,f}$-$(2\pi)^{b,e}$
\\ \hline
 U(1,3,4,8)  & $(2\pi)^{a,d,f}$-$(2\pi)^{b,e,g}$  & U(1,3,5,6)  & $(2\pi)^{a,e}$-$(2\pi)^{b}$ & U(1,3,5,7)  & $(2\pi)^{a,e}$-$(2\pi)^{b,f}$
 \\ \hline
 U(1,3,5,8)  & $(2\pi)^{a,e,g}$-$(2\pi)^{b,f}$  & U(1,3,6,7)  & $(2\pi)^{a,f}$-$(2\pi)^{b}$ & U(1,3,6,8)  & $(2\pi)^{a,f}$-$(2\pi)^{b,g}$
\\ \hline
 U(1,3,7,8)  & $(2\pi)^{a,g}$-$(2\pi)^{b}$  & U(1,4,5,6)  & $(2\pi)^{a,c,e}$-$(2\pi)^{b}$ & U(1,4,5,7)  & $(2\pi)^{a,c,e}$-$(2\pi)^{b,f}$
\\ \hline
 U(1,4,5,8)  & $(2\pi)^{a,c,e,g}$-$(2\pi)^{b,f}$  & U(1,4,6,7)  & $(2\pi)^{a,c,f}$-$(2\pi)^{b}$ & U(1,4,6,8)  & $(2\pi)^{a,c,f}$-
$(2\pi)^{b,g}$  \\ \hline
 U(1,4,7,8)  & $(2\pi)^{a,c,g}$-$(2\pi)^{b}$  & U(1,5,6,7)  & $(2\pi)^{a,c,f}$-$(2\pi)^{b,d}$ & U(1,5,6,8)  & $(2\pi)^{a,c,f}$-
$(2\pi)^{b,d,g}$  \\ \hline
 U(1,5,7,8)  & $(2\pi)^{a,c,g}$-$(2\pi)^{b,d}$  & U(1,6,7,8)  & $(2\pi)^{a,c,e,g}$-$(2\pi)^{b,d}$ &             &             \\ \hline
\end{tabular}  
\end{center}
\end{table}

\begin{figure}
\begin{center}
\epsfig{file=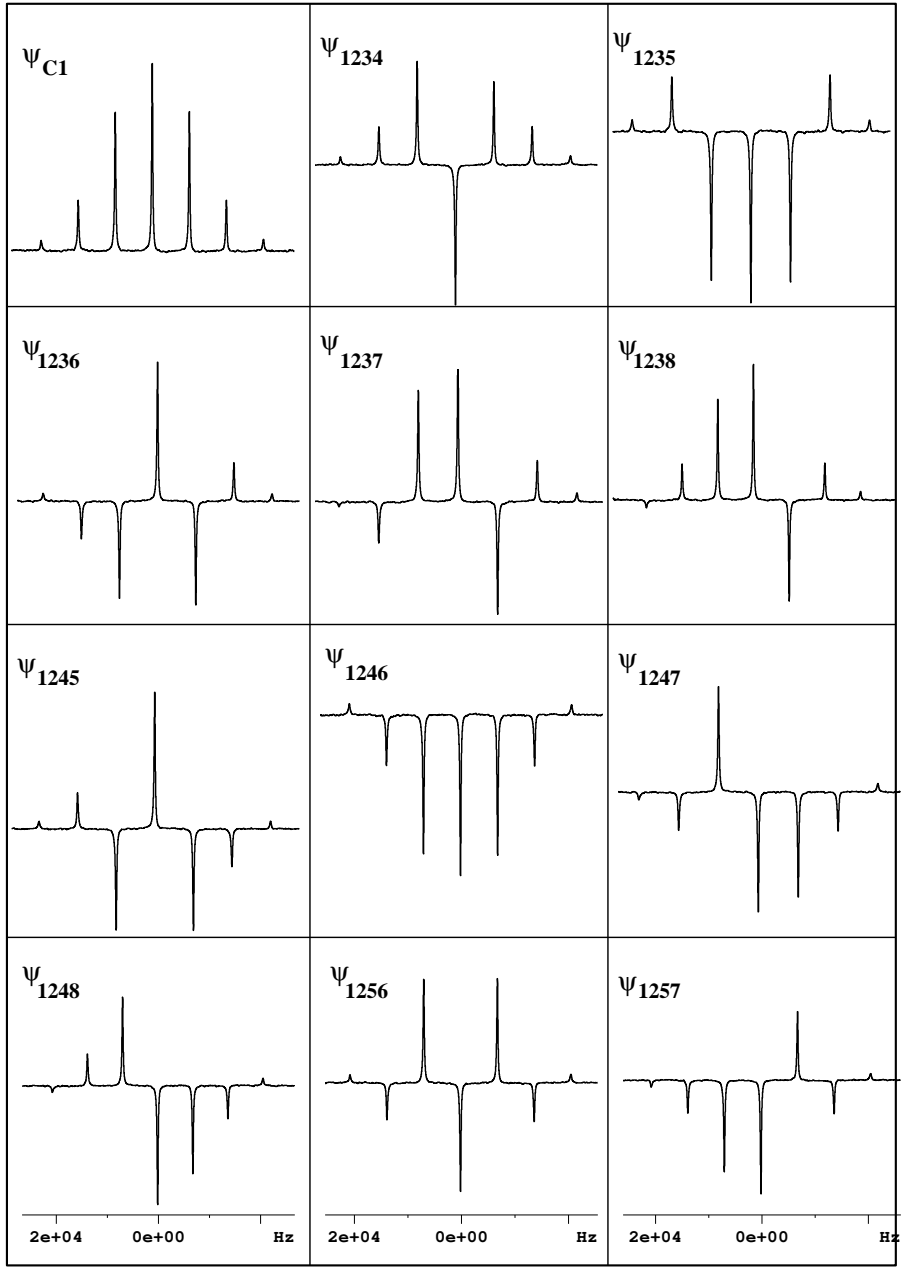,width=12cm,angle=0}
\caption{Implementation of Collins version of 3-qubit DJ algorithm (Fig. \ref{djckt}). Spectrum of $\psi_{c1}$ which corresponds to
a constant function U=I. Spectra of $\psi_{1,k,l,m}$ (Eq. \ref{6eqn10}) which correspond to balanced
functions U(1,k,l,m) (Eq. \ref{6eqn6}).
The required MF pulses for implementing U(1,k,l,m) are given in Table (2). The duration of each MF pulse is 1.425ms.
Each spectrum is recorded in 4 scans}
\label{dj1}
\end{center}
\end{figure}

{\bf (V) Detection:} 
The single quantum transitions (a, b, ..g) of the final states (Eq. \ref{6eqn9},\ref{6eqn10}) are detected. 
Figure (\ref{dj1}) shows the spectrum of $\vert \psi_{c1} \rangle$ and various $\vert \psi_{1,k,l,m} \rangle$. 
From the shape of the spectrum one can conclude, whether the final state represents a constant (or) a balanced function. 
For constant function ($\vert \psi_{c1} \rangle$) none of the peaks
are inverted, whereas for balanced functions ($\vert \psi_{1klm} \rangle$) atleast one of the peaks is inverted. 
Furthermore 
the phases of transitions confirm the final state $\vert \psi_{1,k,l,m} \rangle$. 
For example in the spectrum of $\vert \psi_{1,2,3,4} \rangle$, transition d is negative since the phase difference between the states 
$\vert 4 \rangle$ and 
$\vert 5 \rangle$ is $e^{i\pi}$, while all others have zero phase difference, similarly one can confirm the other states.

\section{4. Conclusions}
Demonstration of quantum computing protocols on various NMR systems, is a promising research area for increasing number of qubits. 
In this work controlled phase shift gates are implemented on a oriented spin-7/2 nucleus, using non-adiabatic geometric phases 
obtained by using selective pulses on single quantum transitions. 
It is also demonstrated here that in an N level system, one can implement a controlled phase shift gate by sandwiching various 
geometric phase shift gates, this method can also be applied to weakly as well as strongly coupled spin-1/2 systems. 
The number of selective pulses are reduced by using multi frequency (MF) pulses. 
Collins version of 3-qubit DJ algorithm is implemented, where the eight basis states are 
collectively treated as a three qubit system. The required controlled phase shift gates of the algorithm, are implemented by using 
MF-$(2\pi)$ pulses.

\section{appendix}

The angular momentum operators of spin-7/2 nucleus, are given by \cite{6ern},

\begin{eqnarray}
&I_z&\vert I,M\rangle=M\vert I,M\rangle \nonumber \\
&I_+&\vert I,M\rangle=\sqrt{(I-M)(I+M+1)}\vert I,(M+1)\rangle \nonumber \\
&I_-&\vert I,M\rangle=\sqrt{(I+M)(I-M+1)}\vert I,(M-1)\rangle,
\end{eqnarray}

where I=7/2 and M= -I, (-I+1),....,+I = -7/2, -5/2, -3/2, -1/2, 1/2, 3/2, 5/2, 7/2.

\begin{eqnarray}
&& \begin{matrix}\vert \frac{7}{2}, \frac{7}{2} \rangle&\vert \frac{7}{2}, \frac{5}{2} \rangle&\vert \frac{7}{2}, \frac{3}{2} \rangle
&\vert \frac{7}{2}, \frac{1}{2} \rangle
&\vert \frac{7}{2}, \frac{-1}{2}\rangle
&\vert \frac{7}{2}, \frac{-3}{2} \rangle&\vert \frac{7}{2}, \frac{-5}{2} \rangle&\vert \frac{7}{2}, \frac{-7}{2} \rangle \end{matrix}
\nonumber \\
I_z=
&&\begin{pmatrix}
\frac{7}{2}&~~~~~0&~~~~~0&~~~~~0&~~~~0&~~~~~~0&~~~~~~~0&~~~~0 \cr
0&~~~~~\frac{5}{2}&~~~~~0&~~~~~0&~~~~0&~~~~~~0&~~~~~~~0&~~~~0 \cr
0&~~~~~0&~~~~~\frac{3}{2}&~~~~~0&~~~~0&~~~~~~0&~~~~~~~0&~~~~0 \cr
0&~~~~~0&~~~~~0&~~~~~\frac{1}{2}&~~~~0&~~~~~~0&~~~~~~~0&~~~~0 \cr
0&~~~~~0&~~~~~0&~~~~~0&~~~~\frac{-1}{2}&~~~~~~0&~~~~~~~0&~~~~0 \cr
0&~~~~~0&~~~~~0&~~~~~0&~~~~0&~~~~~~\frac{-3}{2}&~~~~~~~0&~~~~0 \cr
0&~~~~~0&~~~~~0&~~~~~0&~~~~0&~~~~~~0&~~~~~~~\frac{-5}{2}&~~~~~0 \cr
0&~~~~~0&~~~~~0&~~~~~0&~~~~0&~~~~~~0&~~~~~~~0&~~~~~\frac{-7}{2} \cr
\end{pmatrix} \nonumber
\begin{matrix}\vert \frac{7}{2}, \frac{7}{2} \rangle \cr \vert \frac{7}{2}, \frac{5}{2} \rangle \cr
\vert \frac{7}{2}, \frac{3}{2}\rangle \cr \vert \frac{7}{2}, \frac{1}{2} \rangle \cr
\vert \frac{7}{2}, \frac{-1}{2} \rangle \cr \vert \frac{7}{2}, \frac{-3}{2} \rangle \cr
\vert \frac{7}{2}, \frac{-5}{2} \rangle \cr \vert \frac{7}{2}, \frac{-7}{2} \rangle, \end{matrix} \\ \label{6eqn2}
\end{eqnarray}

\begin{eqnarray}
I_x= \frac{I_++I_-}{2}=\begin{pmatrix}
0&\frac{\sqrt{7}}{2}&&0&0&0&0&0 \cr
\frac{\sqrt{7}}{2}&0&\sqrt{3}&0&0&0&0&0 \cr
0&\sqrt{3}&0&\frac{\sqrt{15}}{2}&0&0&0&0 \cr
0&0&\frac{\sqrt{15}}{2}&0&2&0&0&0 \cr
0&0&0&2&0&\frac{\sqrt{15}}{2}&0&0 \cr
0&0&0&0&\frac{\sqrt{15}}{2}&0&\sqrt{3}&0 \cr
0&0&0&0&0&\sqrt{3}&0&\frac{\sqrt{7}}{2} \cr
0&0&0&0&0&0&\frac{\sqrt{7}}{2}&0
\end{pmatrix},  \label{6eqn3}
\end{eqnarray}

\begin{eqnarray}
I_y= \frac{I_+-I_-}{2i}=\begin{pmatrix}
0&\frac{-i\sqrt{7}}{2}&&0&0&0&0&0 \cr
\frac{i\sqrt{7}}{2}&0&-i\sqrt{3}&0&0&0&0&0 \cr
0&i\sqrt{3}&0&\frac{-i\sqrt{15}}{2}&0&0&0&0 \cr
0&0&\frac{i\sqrt{15}}{2}&0&-2i&0&0&0 \cr
0&0&0&2i&0&\frac{-i\sqrt{15}}{2}&0&0 \cr
0&0&0&0&\frac{i\sqrt{15}}{2}&0&-i\sqrt{3}&0 \cr
0&0&0&0&0&i\sqrt{3}&0&\frac{-i\sqrt{7}}{2} \cr
0&0&0&0&0&0&\frac{i\sqrt{7}}{2}&0
\end{pmatrix}.  \label{6eqn4}
\end{eqnarray}

\pagebreak

\pagebreak

\pagebreak

\pagebreak

\pagebreak

\pagebreak

\pagebreak

\end{document}